# Nanoarchitecture in the black wings of *Troides magellanus*: a natural case of absorption enhancement in photonic materials

Aline Herman*[a], Cédric Vandenbem[a], Olivier Deparis[a], Priscilla Simonis[a], Jean Pol Vigneron[a]
[a]Research Center in Physics of Matter and Radiation (PMR), Facultés Universitaires Notre Dame de la Paix (FUNDP), 61 rue de Bruxelles, 5000 Namur, Belgium

**ABSTRACT**

The birdwings butterfly *Troides magellanus* possesses interesting properties for light and thermal radiation management. The black wings of the male exhibit strong (98%) absorption of visible light as well as two strong absorption peaks in the infrared (3 µm and 6 µm) both due to chitin. These peaks are located in the spectral region where the black body emits at 313K. The study of absorption enhancement in this butterfly could be helpful to design highly absorbent biomimetic materials. Observations of the wings using a scanning electron microscope (SEM) reveal that the scales covering the wings are deeply nanostructured. A periodic three-dimensional (3D) model of the scale nanoarchitecture is elaborated and used for numerical transfer-matrix simulations of the absorption spectrum. The complex refractive index of the wing material is approximated by a multi-oscillator Lorentz model, leading to a broad absorption in the visible range as well as two peaks in the infrared. The absorption peak intensities turn out to be dependent on the complexity of the nanostructures. This result clearly demonstrates a structural effect on the absorption. Finally, a comparison with a planar layer of identical refractive index and material volume lead us to conclude that the absorption is enhanced by nanostructures.

**Keywords:** absorption enhancement, structural absorption, black body emissivity, thermal regulation, nanostructures, butterfly

## 1. INTRODUCTION

The *Troides magellanus* male butterfly (Figure 1) belongs to the birdwings family which comprises most of the biggest butterflies on earth [1]. This family contains several geni: *Ornitoptera*, *Trogonoptera* and *Troides*. The *Troides m.* butterfly displays very contrasted colors: yellow iridescent and fluorescent hind wings and ultrablack forewings. The latter are involved in the thermal radiation management of the butterfly: a negative feedback which stabilizes its body temperature [2]. Indeed, the black wings are extremely absorbing in the visible range of the solar spectrum (98%). The absorbed energy is then used to warm up the muscles before flying [3]. However, heat accumulation is dangerous for the butterfly which has to maintain its body temperature constant, around 40°C [4]. Above this temperature, the butterfly could not fly anymore [5]. In order to keep constant temperature, the butterfly emits infrared radiation as a black body [6]. The black body emission peak down-shifts from 9.9µm to 9.25µm when temperature rises from 20°C to 40°C. It is therefore very useful for the butterfly to enhance its emissivity (absorption) around 10µm. Indeed, absorption peaks located around this wavelength play the role of the above-mentioned negative feedback mechanism. In the *Troides magellanus*, a high visible range absorption combined with high infrared emissivity contribute to efficient thermal regulation.

The desirable properties are obtained thanks to the composition of the black wings. They are made of chitin and melanin. Chitin is the basic constituent of the structure of the wings and is a transparent material. Melanin is the absorbing pigment giving the black color to the wings. Proportion, concentration and form (diluted or not) of melanin into the wings are not known. However, it is well known that melanin and chitin give both rise to broadband absorption of visible wavelengths [7][8][9] as well as two discrete peaks in the infrared region, around 3µm and 6µm [10][11][12].

In addition to pigment, nanostructures present on the black wings are expected to play a role in the absorption and emissivity processes. Indeed, in comparison with a homogenous slab having the same quantity of material, a nanostructured layer shows a higher absorption (as it will be shown here after).

*aline.herman@fundp.ac.be; phone 003281724705

The aim of our study is to determine the influence on absorption/emissivity properties of the different nanostructures found on the scales of the black wings of *Troides magellanus*. First of all, optical as well as Scanning Electron Microscopy (S-E-M) characterizations are carried out. Based on morphological data, four structural models of the wings are elaborated. The resonant material absorption coefficient is described using a multi-oscillator Lorentz model. Absorption spectra of the four wing models are numerically calculated using a 3D-transfer-matrix method. Finally, each model is analyzed in terms of photonic performances (taking into account both the visible absorption and the infrared emissivity).

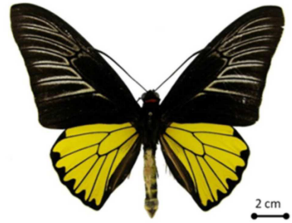

Figure 1. Photography of the *Troides magellanus* butterfly (male).

## 2. OPTICAL CHARACTERIZATION

Reflectance (R) and transmittance (T) spectral measurements are carried out using both Perkin-Elemer UV/Visible Lambda 750S spectrophotometer (equipped with an integrating sphere) and infrared Vertex 70 spectrophotometer. Based on these two measurements, the absorption spectrum is deduced from $A = 1 - T - R$. The reflectance, transmittance and absorption spectra are shown in Figure 2. The absorption spectrum shows a strong broad peak in the visible region (from 400nm to 700nm) with an average value of 98%. In the infrared region, two peaks are observed; the first one around 3µm and the second one around 6µm. These peaks originate from the chemical composition of the biological material (melanin and chitin) forming the wing. More specifically, they arise from the vibration modes of amides (I and II) located around 1626 - 1560cm$^{-1}$ and at 3000cm$^{-1}$ [10][11][12].

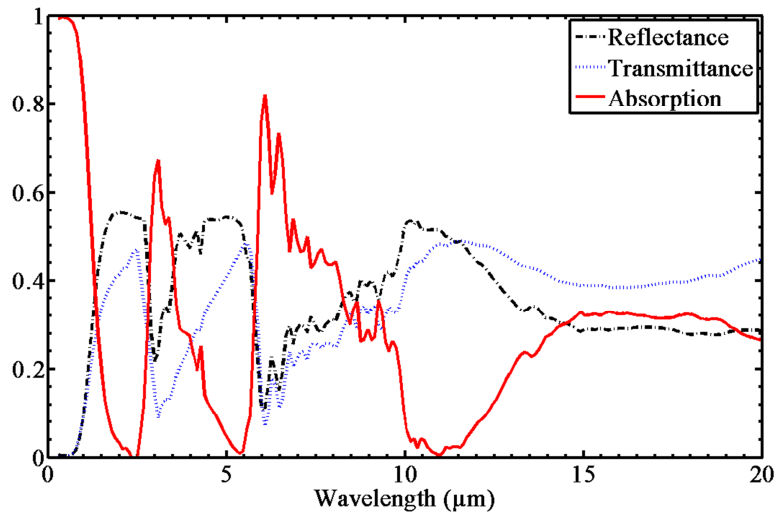

Figure 2. Reflectance (dashed-dotted line), transmittance (dotted line) and absorption (solid line) experimental spectra of the black wings of the *Troides magellanus* butterfly.

## 3. MORPHOLOGY AND MODELS

### 3.1 Scanning electron microscopy observations

A sample taken from one of the black wings of a *Troides magellanus* specimen (Figure 1) is observed using a LEICA S 440 Scanning Electron Microscope (SEM). Images taken at low magnification show that the wing is covered by scales with average dimensions of $189 \mu m \times 70 \mu m \times 4.7 \mu m$ (Figure 3a). A typical scale is shown in Figure 3b. Each scale possesses two membranes: a lower flat membrane and an upper nanostructured one. The latter consists in various nanostructures, forming a complex nanoarchitecture (Figure 3c). The nanoarchitecture is formed by five principal elements: a roof-like structure (1) which is covered by tilted ridges (2), holes (3) which are present in the spacers (4) between two consecutive roof bases and pillars (5) linking the upper membrane to the lower one. Each of these elements has wavelength-scale dimensions and thus could influence light propagation and absorption.

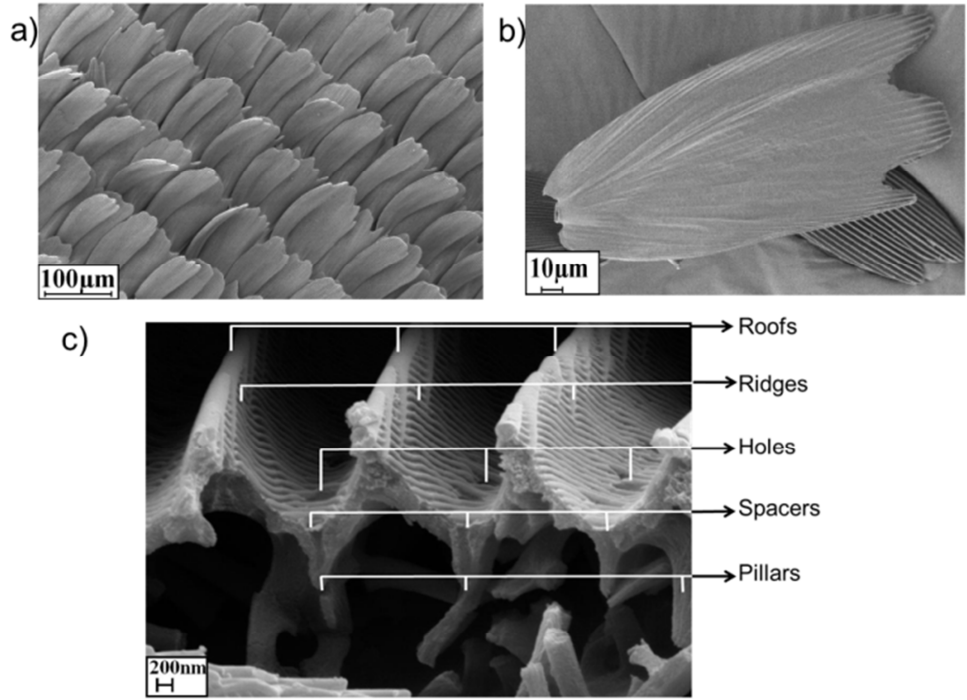

Figure 3. a) SEM image of the scales covering the black wings of *Troides magellanus*, b) SEM image of one scale, c) nanoarchitecture of a scale: five types of nanostructures are identified.

### 3.2 Nanoarchitecture models

From SEM images, a periodic three-dimensional (3D) model of the scale architecture is established and used in numerical simulations in order to calculate the absorption spectrum by means of a 3D-transfer-matrix electromagnetic method. In order to study the influence of each element of the nanoarchitecture on the photonic response, the 3D model is built step by step. First of all, only the roof-spacer 2D structure is considered (Figure 4a). Several elements are progressively added to the roof-spacer structure to form three other (3D) models. To form the second model (Figure 4b), pillars and inferior membrane are added to the first model. Then, the tilted ridges are included to the 2nd model in order to form the 3nd model (Figure 4c). The last model (Figure 4d) includes the five elements identified in Figure 3. Note that each model is formed by assembling rectangular blocks. The use of homogeneous rectangular blocks is required for applying the transfer-matrix method. Parameters used for the numerical simulations are listed in Table 1.

Table 1. Parameters used for the modeling of the nanoarchitecture present on the black wings of the *Troides magellanus* butterfly.

| Parameter | Value |
|---|---|
| Membrane thickness (nm) | 250 |
| Roof base (nm) | 100 |
| Roof height (nm) | 1200 |
| Gap between the membranes (nm) | 1500 |
| Period of roof-spacer structure (nm) | 1500 |
| Pillar thickness (nm) | 250 |
| Hole radius (nm) | 500 |
| Ridge thickness (nm) | 50 |
| Gap between ridges (nm) | 200 |
| Ridge tilt angle (°) | 60 |

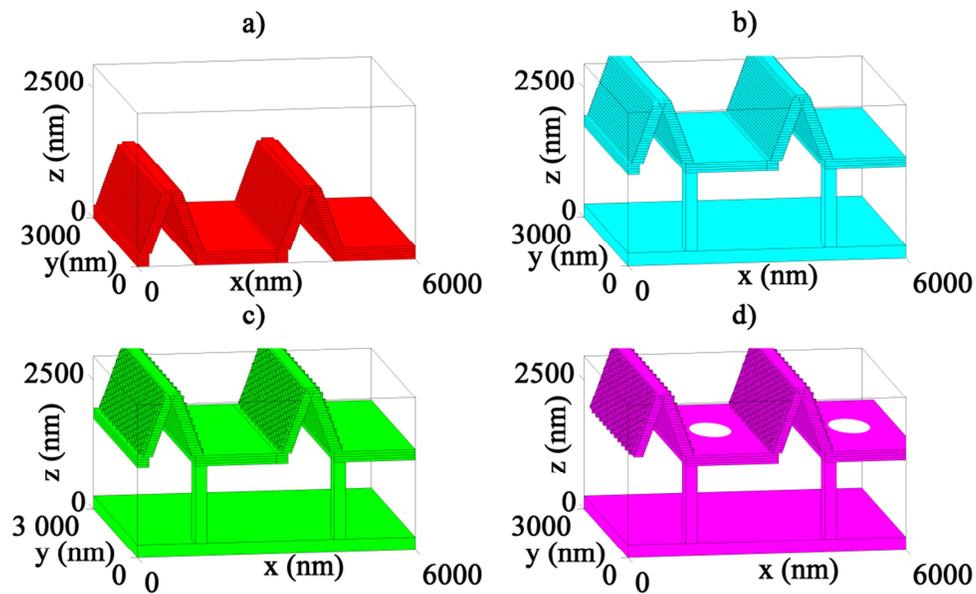

Figure 4. Three dimensional (3D) views of the models representing the nanoarchitecture formed on the scales of the *Troides magellanus a*) model 1 (2D) results from the periodic repetition of the roof-spacer like structure; b) model 2 (3D) includes the inferior membrane and the pillars; c) tilted ridges (60°) are added to create model 3 (3D); d) model 4 (3D) is the most complete one and takes in account all the elements which are observed on Figure 3c.

# 4. MODELING OF THE RESONANT MATERIAL ABSORPTION COEFFICIENT

Numerical simulations of the absorption spectrum require the knowledge of the resonant absorption coefficient of the wing, α. This coefficient is related to the complex refractive index $n+ik$, via $k = \dfrac{\alpha \lambda}{4\pi}$. The exact determination of $n(\lambda)$ and $k(\lambda)$ is extremely difficult because the concentration of melanin, the absorbing pigment present in the black wings, is not known. However, the absorption coefficient of materials containing melanin depends on the melanin concentration [8] [13]. Moreover, ellipsometry measurements are not possible because of the nature of the sample (thin, structured and inhomogeneous). To bypass this difficulty, the spectral dependence of the complex permittivity $\varepsilon = (n+ik)^2$, is modeled using a multi-oscillator Lorentz dispersion law:

$$\varepsilon(\omega) = \varepsilon_\infty + \sum_j \frac{\omega_{p,j}^2}{\omega_{R,j}^2 - \omega^2 - i\gamma_j \omega} \qquad (1)$$

where $\varepsilon(\omega)$ is the wavelength-dependent complex permittivity, $\varepsilon_\infty$ is the contribution of higher energy resonances to the real part of $\varepsilon$, $\omega_{p,j}$ is the plasma frequency of oscillator number $j$, $\omega_{R,j}$ is the resonance frequency of oscillator number $j$ and $\gamma_j$ is the resonance damping coefficient for the $j^{th}$ resonance. Six Lorentz oscillators ($j = 1;6$) are used in the present case. The positions of the resonance wavelengths are chosen according to the literature. Indeed, melanin gives rise to a broadband absorption in the visible [7] [8] [9] and chitin gives rise to two infrared peaks (3µm and 6µm) [10] [11] [12]. The parameters of the Lorentz model are chosen considering the fact that the real part of the refractive index, $n$, has to be, on average, close to $1.56$, i.e. the commonly used value for chitin in the visible-near infrared range. Furthermore, because the imaginary part of the refractive index $k$ is related to the absorption coefficient α, we selecte resonance frequencies $\omega_{R,j}$ (up to six different ones) in order to match as close as possible the peak positions and widths which are found in the experimentally measured absorption spectrum (Figure 2). The parameters of each oscillator are given in Table 2. The resulting spectra of $n$ and $k$ are shown in Figure 5.

Table 2. Parameters of the six Lorentz oscillators used to model the complex refractive index of the biological material (chitin with melanin). The symbol $O_i$ represents the $i^{th}$ oscillator. $\varepsilon_\infty$ is equal to 2.2.

|  | $O_1$ | $O_2$ | $O_3$ | $O_4$ | $O_5$ | $O_6$ |
|---|---|---|---|---|---|---|
| $\lambda_P$ (nm) | 412.5 | 832 | 2030 | 8400 | 16000 | 15300 |
| $\lambda_R$ (nm) | 330 | 520 | 700 | 3000 | 6400 | 17000 |
| γ (Hz) | $1.7\times10^{15}$ | $0.9\times10^{15}$ | $0.3\times10^{15}$ | $0.3\times10^{15}$ | $0.01\times10^{15}$ | $0.02\times10^{15}$ |

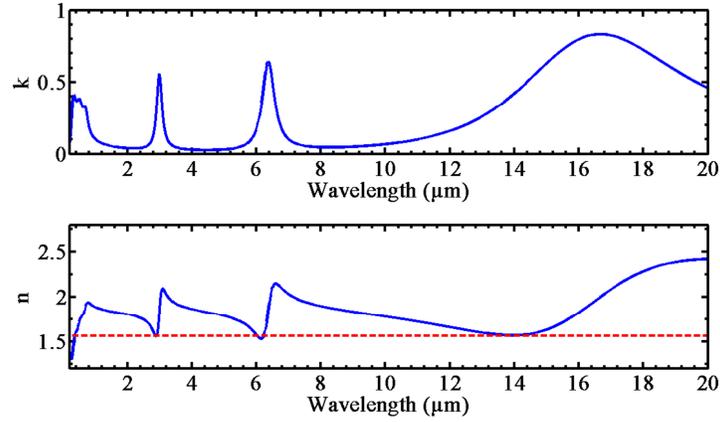

Figure 5. Spectra of $n$ and $k$, calculated using a 6-oscillators Lorentz model. These spectral data are used to approximate the wavelength-dependent complex permittivity of the black wing, $\varepsilon(\omega) = [n(\omega)+ik(\omega)]^2$. The dashed line on the bottom graph is the commonly used average real refractive index of chitin ($n = 1.56$).

## 5. ELECTROMAGNETIC CALCULATION METHOD

The goal of the study is to determine the influence of each nanostructure on the absorption and emissivity properties. For this purpose, the absorption spectrum of each model is simulated numerically using a three-dimensional transfer- matrix electromagnetic method. In order to apply this method, the structure has to be divided into layers which are parallel to the lower flat membrane plane. The transfer matrix method could be applied for both homogeneous and inhomogeneous photonic media. For the latter case, the permittivity (dielectric function $\varepsilon$) has to be periodic in the lateral directions ($x$ and $y$ axes) but could vary arbitrarily in the vertical direction ($z$ axis) normal to the plane of the layers. The periodic variations of $\varepsilon$ are represented by means of Fourier series.

The problem of electromagnetic wave propagation in the inhomogeneous multilayer structure is treated by solving rigorously Maxwell's equations. The amplitude of the electromagnetic field on one side of a layer is connected to the amplitude at the other side. The transfer matrix (TM) of the layer is then converted into a scattering matrix (SM) for numerical stability reasons. The SM of the layers are assembled using Pendry's formula in order to determine the SM of the multilayer structure [14]. This matrix is then used to calculate the amplitudes of scattered fields in all diffraction directions. From the scattered field amplitudes, the energy fluxes are calculated in incidence and emergence media and lead to the reflectance and the transmittance, respectively.

The method could be used with both constant refractive indexes (independent of the wavelength) and wavelength-dependent refractive indexes. The complex refractive index, with real part noted $n$ and imaginary part noted $k$, is related to the complex permittivity $\varepsilon$ by $n+ik = \sqrt{\varepsilon}$.

The hemispherical (total) reflectance (R) and transmittance (T) are calculated for each model and used to determine the absorption ($A = 1-T-R$). In order to isolate the influence of each nanostructure on the absorption spectrum, the absorption of each model (Figure 4) is compared with the absorption of a planar homogeneous slab (without any structures) containing the same volume of absorbing material.

## 6. SIMULATION RESULTS

The photonic response of a nanoarchitecture model arises from both the nanostructures and the extinction coefficient of the material ($k$). In order to isolate the influence of the nanostructures, we use a uniform slab with identical refractive index and identical volume of absorbing material than the model, as a point of comparison. Therefore each model has its own reference.

In all simulations, we use non-polarized light at normal incidence where ($x,z$) plane is the plane of incidence (polar and azimutal angles equal to $0°$). Numerical convergence of the results is tested before. For the 2D model (model 1), 21 plane waves are used in the $x$ direction and only one plane wave along the $y$ axis. For the other models, three-dimensional ones, 21 ($x$ direction)×7 ($y$ direction) plane waves in the Fourier expansions turn out to be necessary to reach convergence. The choice of the number of plane waves depends on the model geometry. The stronger the spatial variations of the structure along a specific direction, the higher the number of plane waves needed in the Fourier expansions in that direction.

### 6.1 Limitation on the absorption by a homogeneous slab

As explained above, we choose homogeneous slabs as references. To get an idea of the dependency of the absorption on $k$, we first perform simulations using various $k$ spectra. The homogeneous slab in the following example has a thickness of 400nm. Figure 6 shows absorption spectra for two different $k$ values. We note that there is a saturation of the maximum value of the absorption for a planar slab (Figure 6c). Even if the peak value of $k$ is increased, absorption does not exceed a maximum value. Therefore, it is not possible to overcome this limit in the absorption by keeping a planar geometry.

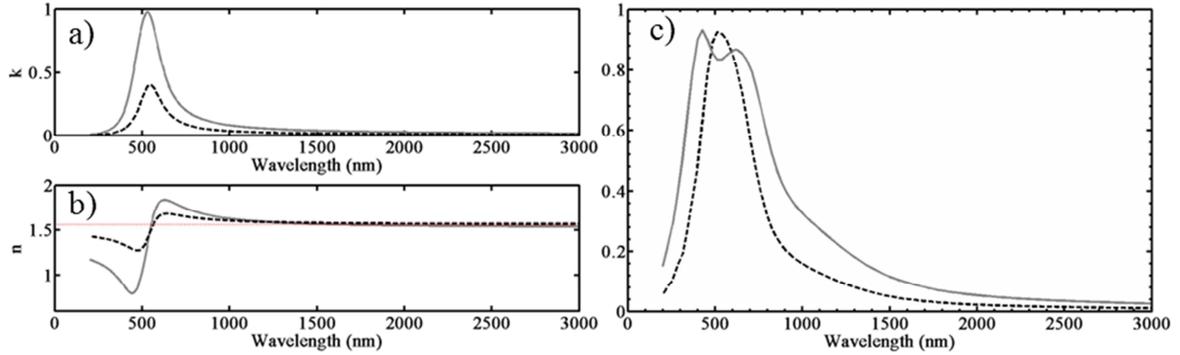

Figure 6. a) Spectra of $k$ used to calculate the absorption spectrum of a homogeneous slab, for two different profiles of $k$; b) Spectra of $n$ used to calculate the absorption spectrum of a homogeneous slab. The dotted line represents the commonly used value of the real part of the refractive index of chitin ($n=1.56$); c) Absorption spectra of a homogenous slab of thickness 400nm. The parameters are the following: dashed dark lines $\lambda_R = 550nm$; $\lambda_P = 929.5nm$; $\varepsilon_\infty = 2.10$; $\gamma = 0.5\times10^{15} Hz$, solid gray lines $\lambda_R = 550nm$; $\lambda_P = 599.5nm$; $\varepsilon_\infty = 1.50$; $\gamma = 0.5\times10^{15} Hz$. Note that $n$ and $k$ are obtained by using a single Lorentz oscillator model.

## 6.2 Absorption spectra of the models

Several absorption spectra are numerically calculated; from the simplest corrugated slab (2D model 1) to the most complex structured slab (3D model 4, full wing). The spectrum of each model is compared with the spectrum of its own reference, i.e. a homogeneous slab with the identical complex refractive and identical volume of material (Figure 7). By nanostructuring the slab, the maximum value of the absorption that is reached for the homogeneous slab can be overcome. The absorption enhancement is more pronounced around the resonance wavelengths of the material refractive index.

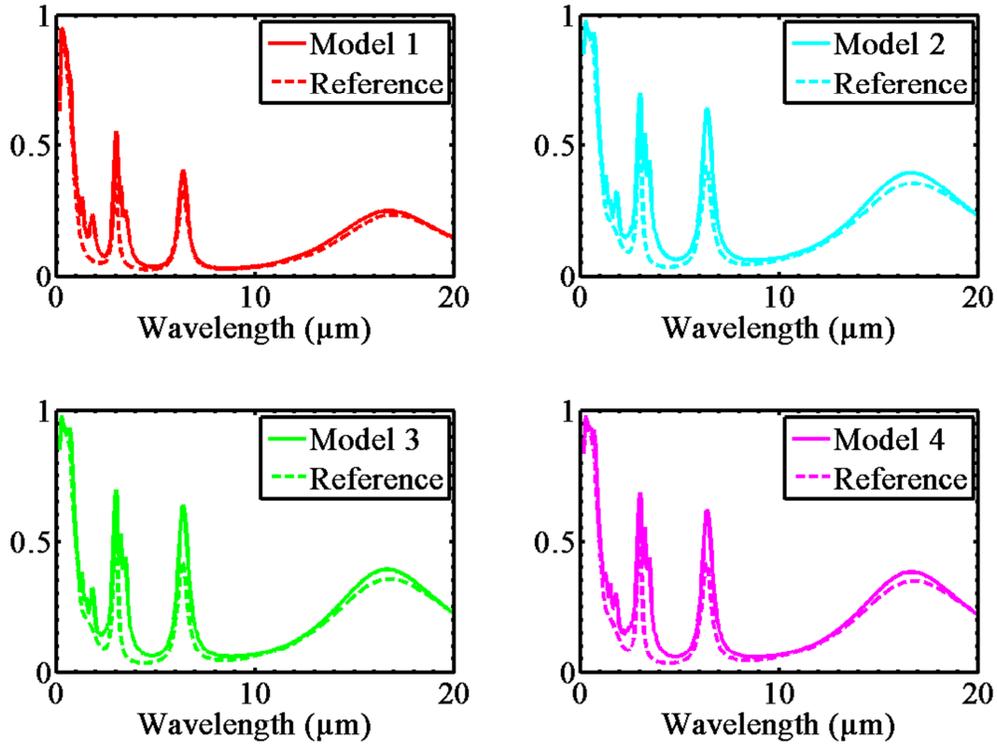

Figure 7. Numerically simulated absorption spectra of the four models (solid lines) defined in Figure 4. Each model is compared with a planar homogeneous slab (dashed lines) with identical volume of material. The thicknesses of the reference slabs are the following: for model 1, 500nm; for model 2, 510nm; for model 3, 512nm and for model 4, 490nm.

# 7. ANALYSIS OF THE RESULTS

The black wings of the *Troides magellanus* play a role in its body thermal regulation. They absorb the major part of the solar spectrum and emit infrared radiation as a black body. Let us discuss now the influence of each nanostructure on this thermal management.

## 7.1 Solar absorption enhancement factor

In order to quantify the effect of the nanostructures on the absorption of the solar light spectrum, we define a solar absorption enhancement factor.

$$A = \frac{\alpha_{model} - \alpha_{reference}}{\alpha_{reference}} \quad (2)$$

where

$$\alpha = \frac{\int AM_{1.5}(\lambda) A(\lambda)\, d\lambda}{\int AM_{1.5}(\lambda)\, d\lambda} \quad (3)$$

and $AM_{1.5}(\lambda)$ is the normalized solar emission spectrum ($AM_{1.5}$ standard) and $A(\lambda)$ is the absorption spectrum of the model or the reference (Figure 8a). The integrals in (3) are carried out over the whole spectral range of the simulations (200nm – 20µm). However, the solar emission spectrum is located in the visible-near infrared range (300nm – 1200nm).

## 7.2 Infrared emissivity enhancement factor

Regarding the infrared emissivity at a given temperature T, we define an emissivity enhancement factor:

$$E(T) = \frac{\varepsilon_{model}(T) - \varepsilon_{reference}(T)}{\varepsilon_{reference}(T)} \quad (4)$$

where

$$\varepsilon(T) = \frac{\int \varepsilon_{black\,body}(\lambda, T)\ A(\lambda) d\lambda}{\int \varepsilon_{black\,body}(\lambda, T) d\lambda} \quad (5)$$

and $\varepsilon_{black\,body}(\lambda, T)$ is the black body emissivity at a given temperature T and $A(\lambda)$ is the absorption spectrum of the model or the reference (Figure 8b). The integrals in (5) are carried out over the whole spectral range of the simulations (200nm – 20µm). However, the black body spectrum is located around 9µm for a temperature of 40°C.

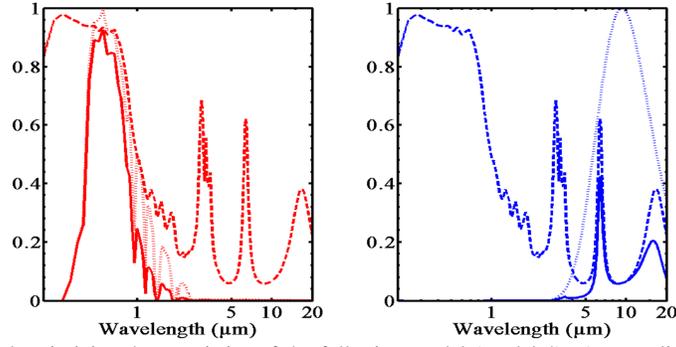

Figure 8. Absorption and emissivity characteristics of the full wing model (model 4); a) normalized ($AM_{1.5}$ standard) solar emission spectrum (dotted-line), absorption spectrum of model 4 (dashed line), product of the spectra $AM_{1.5}(\lambda)A(\lambda)$ (solid line); b) normalized black body emissivity at 40°C (dotted-line), absorption spectrum of model 4 (dashed line), product of the spectra $\varepsilon_{black\,body}(\lambda,T)\,A(\lambda)$ (solid line). Note that the $x$ axis is in logarithmic scale.

### 7.3 Analysis of the performance

In order to identify the most performing model among the four ones, we calculate the enhancement factors $A$ and $E(T)$ for each model (Table 3).

Table 3. Values of the solar absorption enhancement factor $A$ and the emissivity enhancement factor $E$ at 20°C and 40°C, for the four investigated models.

|  | Thickness of the reference slabs (nm) | A (%) | $E(T = 20°C)$ (%) | $E(T = 40°C)$ (%) |
|---|---|---|---|---|
| **Model 1** | 500 | 13 | 10 | 11 |
| **Model 2** | 510 | 8 | 17 | 18 |
| **Model 3** | 512 | 8 | 15 | 17 |
| **Model 4** | 490 | 9 | 14 | 16 |

For all models, the values of $A$ are positive: all the nanostructured models are more absorbing than their respective reference. If we classify the models from the most absorbing one to the less absorbing one, we find that the best model is the most simple one (model 1), the complete model arriving in second position, followed by model 2 and model 3.

Regarding the emissivity performance, results are a bit different. The values of $E$ are always positive, confirming the fact that the nanostructured models are more performing than a planar slab with the same volume of material. If the temperature is increased from 20°C to 40°C, the values of $E$ are increased as well. The higher the temperature, the bigger the emissivity. This increase of the infrared radiation emission with temperature helps to keep the body temperature of the butterfly close to its optimal value. As before, we can classify the models from the most emitting one to the less emitting one. The result is: model 2, model 3, model 4 and model 1. Therefore, the most absorbing model is the less emitting one and conversely.

The most simple model (model 1) has the highest solar absorption and, at the contrary, the lowest emissivity. From the point of view of thermal management, the second model absorbs the major part of the solar radiation and emits as much as possible in the infrared. Therefore, model 2 gives the best performance from this point of view. The most structured model (model 4) which is also the closest to the real scale structure of the *Troides magellanus*, is the one which uses the smallest quantity of material (Table 3), but achieves good performances, close to the best one (model 2).

The fact that the most structured model gives almost the best performances with a minimum of material suggests that the scale structure of the *Troides magellanus* has evolved in such a way to achieve thermal body regulation with the lowest weight of the wings. Although this interpretation is always questionable from a biological point of view, it is certainly an interesting guide line for designing highly absorbing/emitting biomimetic nanostructured materials. Indeed, the lesson to be learnt from the *Troides magellanus* case study is that it is possible, at the same time, to have both good absorption and emissivity and to minimize the weight of the nanostructured material.

## 8. CONCLUSIONS

In order to accumulate enough energy for warming up their muscles before flying, butterflies such as *Troides magellanus* need to maximize their solar absorption (visible wavelength range). As a result of this strong absorption, its body temperature increases and this heating could be dangerous for it. In order to maintain a constant body temperature, around 40°C, the butterfly must evacuate heat through thermal radiation emission like a black body (infrared emissivity).

In this study, the origin of the thermal management taking place in the black wings of the butterfly *Troides magellanus* is elucidated. Based on three-dimensional transfer-matrix numerical simulations of the absorption spectrum of realistic wing scale models, we conclude that the absorption/emissivity is not only due to melanin pigments in the bulk material but also depend on the type of nanostructures covering the scales of the wing.

Several models are built by increasing the complexity of the nanostructure. The simplest model is a two-dimensional roof-spacer-like structure. The most complex three-dimensional model, which is the closest to the real wing structure, contains an inferior membrane, holes in the spacers and tilted ridges on the roofs. Comparison of the absorption spectrum of each structural model with the absorption of a plane homogeneous slab of identical volume of absorbing material clearly shows that the absorption is not only due to pigments but also to a structural effect. Compared to a planar slab, the nanostructures of the black wings enhance the solar absorption by $\sim 10\%$ and the infrared emissivity at 40°C by $\sim 17\%$.

Nevertheless, there remain some discrepancies between the experimental absorption spectrum and the simulated one, even for the most realistic model. They are thought to be due to the fact that the complex refractive index is approximated by a Lorentz oscillator model, which fits reasonably the experimental absorption data. Another explanation is that we use a perfectly ordered (periodic) model, although, in reality, the nanostructures exhibit some disorder.

It also turns out that the most absorbing structure (in the visible range) is not the most performing one in terms of infrared emissivity and reversely. However, optimization of both absorption and emissivity is found to be possible with the most complex model, at the expense of a slight decrease of the absorption enhancement factor. Interestingly, this model is the closest to the real structure of the wing sales of the *Troides magellanus* butterfly.

## ACKNOWLEDGMENTS

J. F. Colomer and E. Van Hooijdonk are acknowledged for SEM images of the sample. This research used resources of the Interuniversity Scientific Computing Facility located at the University of Namur, Belgium, which is supported by the F.R.S.-FNRS under convention No. 2.4617.07.